\documentclass[letterpaper,journal]{IEEEtran}
\usepackage{amsmath,amsfonts}
\usepackage{algorithmic}
\usepackage{algorithm}
\usepackage{array}
\usepackage[caption=false,font=normalsize,labelfont=sf,textfont=sf]{subfig}
\usepackage{textcomp}
\usepackage{stfloats}
\usepackage{url}
\usepackage{verbatim}
\usepackage{graphicx}
\usepackage{cite}
\usepackage{newtxtext,newtxmath}  
\usepackage{xcolor}
\usepackage{soul}   
\soulregister{\cite}{1}

\begin{document}

\title{Dynamic Beam Shaping Using a Wavelength-Adaptive Diffractive Neural Network for Laser-Assisted Manufacturing}

\author{Bharathy Jacob, John Rozario Jegaraj, Nithyanandan Kanagaraj

\thanks{Bharathy Jacob is a Research Scholar at the Ultrafast Fiber Optics \& Smart Photonic Technologies Lab, Department of Physics, Indian Institute of Technology Hyderabad, India – 502284.}

\thanks{John Rozario Jegaraj is Sc F, associated with the Defence Research \& Development Laboratory, Kanchanbagh, Hyderabad, Telangana, India – 500005.}

\thanks{\textbf{Corresponding author}: Nithyanandan Kanagaraj is an Assistant Professor and leads the Ultrafast Fiber Optics \& Smart Photonic Technologies Lab, Department of Physics, Indian Institute of Technology Hyderabad, India – 502284. (email: nithyan@phy.iith.ac.in)}}





\maketitle
\begin{abstract}

Laser-based manufacturing has emerged as a promising alternative to conventional thermal and mechanical processing owing to its precision, versatility, and ability to work across diverse materials. In particular, tailoring the spatial intensity distribution of laser beams on the fly is pivotal for ensuring keyhole stability, minimizing defects, and enhancing processing quality. To address this need, we propose a multifunctional optical platform designed through a Diffractive Neural Network (DNN) that provides wavelength adaptability for three industrially relevant wavelengths—915 nm, 1064 nm, and 1550 nm—while dynamically generating distinct beam profiles at specified propagation planes. The proposed platform not only enables static beam shaping but also supports dynamic beam engineering, including programmable sequencing between profiles, which is highly desirable for optimal manufacturing solutions. With its multifunctionality and adaptability, the DNN-based architecture establishes a transformative pathway for next-generation laser manufacturing, aligning with the industrial revolution while unlocking opportunities in biomedical optics, free-space communications, and sensing, etc.,

\end{abstract}


\section{Introduction}
From fabricating tiny circuits in microelectronics to assembling the frames of electric vehicles, lasers have steadily transformed how we build the world around us~\cite{dunsky2005laser,demir2023challenges}. The incredible ability of lasers to cut, weld, and shape materials with high precision—without ever touching the surface—has made them indispensable across industries like aerospace, automotive, and medical device manufacturing~\cite{bogue2015lasers}. As the demand for smaller, lighter, and more sustainable products grows, traditional tools often fall short to meet the required precision and accuracy. Lasers, however, adapt with ease, offering speed, accuracy, and minimal waste, thereby paving the way for sustainable manufacturing~\cite{gopal2023laser}. Its role is particularly critical in high-tech fields like electric vehicle manufacturing, where precision and scalability are paramount~\cite{demir2023challenges}, emphasizing lasers as a driving enabler for the future of advanced manufacturing~\cite{kruger2013uv}.

One notable example of this technological shift is laser welding, which is increasingly replacing traditional mechanical machining and thermal welding to achieve greater precision, efficiency, and control ~\cite{xiao2022dissimilar}. Conventional methods often require substantial heat input, leading to wide, shallow welds and significant thermal distortion—especially problematic in thin or heat-sensitive components. Laser welding, by contrast, focuses high energy density into a small spot, enabling deeper penetration with a smaller heat-affected zone ~\cite{saediardahaei2024comparative}. This precision is achieved through the formation of a vapor-filled cavity known as a keyhole, which allows high-aspect-ratio welds that improve joint strength and reduce distortion, particularly in materials like steel and nickel. However, keyhole instability can arise in reflective or low-absorption metals such as copper and aluminum, sometimes resulting in defects like porosity and spatter~\cite{wang2023keyhole, saediardahaei2024toward,chopde2017study}.

Despite these challenges, laser-based materials processing remains a preferred choice due to its precision, tunability, and adaptability. Although some drawbacks can be mitigated by core parameters such as wavelength, pulse energy, pulse duration, repetition rate, and feed rate ~\cite{hribar2022influence, legall2019influence, rucker2022unwanted, holland2022influence, freitag2018influence}, a major turnaround in recent years has been the growing emphasis on laser beam engineering. This approach enables precise control over laser–material interactions ~\cite{duocastella2012bessel}, significantly enhancing the quality and specificity of outcomes. For instance, top-hat beams offer a uniform energy distribution that improves surface quality, reduces taper angles, and improves edge definition, outperforming Gaussian beams in microstructuring applications ~\cite{le2020effects, homburg2008laser}. Similarly, Bessel beams are well suited for processing uneven surfaces and enabling high-aspect-ratio features ~\cite{duocastella2012bessel, siems2019beam}, whereas annular beams are advantageous in micromachining, deep-hole drilling, and high-resolution multiphoton polymerization ~\cite{duocastella2012bessel}. Although these static beam profiles have proven to be highly effective, they remain fixed during processing, limiting their adaptability. In contrast, dynamic beam engineering ~\cite{grunewald2022flexible, belay2022dynamic} is emerging as a transformative solution to modern manufacturing needs.

Dynamic beam engineering ~\cite{grubert2024enhancing,putsch2016adaptive,belay2022dynamic} refers to the ability to actively and rapidly modify the laser intensity profile on-the-fly. Unlike static shaping, which locks the process to a single-beam distribution, dynamic methods allow the beam profile to evolve continuously during processing, enabling adaptive and optimized interactions with the material. This capability not only improves surface finish, defect suppression, and overall processing efficiency but also opens the door to highly flexible, task-specific laser strategies. By tailoring the beam response to the instantaneous state of the material, dynamic beam engineering could potentially emerge as a cornerstone of next-generation laser manufacturing, bridging the gap between fundamental laser physics and application-driven performance.

One particularly powerful technique within this paradigm of dynamic beam engineering is beam shape sequencing. Rather than relying on a single beam profile, this approach involves cycling through a sequence of pre-defined shapes, each engineered for a specific purpose. For example, one shape may stabilize the keyhole and reduce spatter during welding, while another may minimize cracking. By executing these shape transitions in rapid succession within microseconds—the laser can meet multiple process objectives in a single pass. Beyond defect control, shape sequencing enables spatially adaptive processing. For example, as the beam moves across different material layers—such as transitioning from a surface coating to an underlying substrate—the sequence can be adjusted to suit each material’s characteristics. This layer-specific tailoring enhances consistency and quality, particularly in complex multi-layer heterogeneous geometries or multi-material assemblies.

Conventional beam shaping techniques typically rely on complex and often bulky optical components such as spatial light modulators (SLM), diffractive optical elements (DOE), refractive beam shapers ~\cite{laskin2012refractive} and axicons ~\cite{dickey2003laser}, to name a few. While these components can be highly effective, they are often limited to narrow wavelength bands due to material dispersion, or constrained to fixed beam profiles, offering minimal adaptability for dynamic or multi-wavelength operations~\cite{brunne2011adaptive}. In many advanced optical applications—ranging from laser-based manufacturing and biomedical imaging to optical trapping and spectroscopy—the ability to shape beams across multiple wavelengths using a single system offers significant advantages. 

Over the years, several approaches have been developed to extend spectral compatibility and broaden the scope of beam-shaping systems. These include hybrid refractive-diffractive systems~\cite{seldowitz1987synthesis}, layered structures composed of complementary materials ~\cite{arieli1998design}, and thickness-engineered ~\cite{sweeney1995harmonic} DOEs that mitigate chromatic dispersion while preserving phase control. Computational strategies like Gerchberg–Saxton (GS) optimization ~\cite{noach1996integrated} and Direct Binary Search (DBS) ~\cite{kim2012design} have also been employed to design phase masks for multi-wavelength beam shaping tasks. However, such approaches generally produce a fixed intensity pattern at a predefined focal plane and lack the ability to dynamically adjust the beam profile.

This gap highlights the need for a fundamentally different strategy—one capable of simultaneously supporting multiple wavelengths while dynamically generating distinct beam profiles at programmable axial planes. Such a system would represent a critical enabler for the next generation of high-precision, application-specific manufacturing and beyond. In this context, we explore the potential of diffractive neural networks (DNNs)—a class of AI-driven optical systems that leverage deep learning for spatial light modulation. First introduced by Lin et al. in 2018 ~\cite{lin2018all}, DNNs consist of multiple phase-modulating layers that manipulate light via diffraction to perform predefined optical functions. During the training phase, these layers are optimized using standard deep learning algorithms to perform tasks such as beam shaping ~\cite{buske2022advanced}, computational imaging ~\cite{mengu2022all}, classification ~\cite{feng2023multi}, or mode sorting ~\cite{zheng2022orthogonality}. Once fabricated, the DNN passively processes optical information as light propagates through it. Unlike conventional optical systems, DNNs offer parallelism, and inference at the speed of light, all while consuming minimal power. This makes them particularly attractive for scenarios demanding real-time, high-throughput, and wavelength-flexible operation.

While early DNN demonstrations were limited to single-wavelength sources and static optical tasks, the field has rapidly evolved to exploit additional degrees of freedom in light, such as polarization and wavelength, enabling multitask optical processing. For example, Luo et al. developed a metasurface-based DNN ~\cite{luo2022metasurface} that performs a parallel classification of different object categories using polarization multiplexing, while Duan et al. implemented a multiwavelength diffractive network (D2NN) ~\cite{duan2023optical} to enhance simultaneous classification performance across different datasets. These advancements underscore the growing ability of DNNs to handle multiple functions within a single, compact optical platform—overcoming the rigidity of traditional diffractive optics.

Our system builds upon these recent developments by first devising a diffractive neural network (DNN) to generate a ring-shaped beam at a predefined target plane for three industrially significant wavelengths. This initial demonstration established the DNN’s ability to handle multi-wavelength beam shaping within a single passive optical system. Building on this capability, we extended the system to perform dynamic beam engineering to generate multiple spatial profiles—namely Gaussian, ring, and top-hat beams—at distinct axial positions for each of the three wavelengths. This represents a significant step forward, not only in terms of compact, passive optical hardware, but also in laser beam engineering for multi-material and multi-layer manufacturing. By combining wavelength- and depth-dependent shaping into a single diffractive network, our approach allows users to switch beam functions in real time without active optics, making it a powerful tool for next-generation laser-based manufacturing systems.

The article is organized as follows: Section II details the numerical design of the proposed DNN architecture. Section III discusses the beam-shaping functionalities demonstrated by the system. Finally, Section IV presents a summary of the key findings and concluding remarks.

\section{Design of the WD-DNN Architecture}
\subsection{Architecture and Operating Principle}

We propose a multifunctional optical system based on DNN with beam-shaping capabilities across multiple operational wavelengths. Unlike conventional artificial neural network, DNN achieve inherent parallelism by utilizing photons instead of electrons. This parallelism can be leveraged through various optical degrees of freedom, including wavelength, polarization, and orbital angular momentum of the source. In this work, we specifically, exploit the wavelength degree of freedom to endow the device with spectral adaptability. Each wavelength channel \(\lambda_i\) (for \(i = 1, 2, \ldots, N\)) can be assigned a unique function or, alternatively, the same function can be replicated across multiple wavelengths. While this redundancy may seem unnecessary, it is in fact essential for ensuring consistent performance when the system is illuminated by sources of different wavelengths. Thus making it a versatile system that meets on-demand beam shaping requirements. In practical applications—particularly those involving systems with wavelength variability—it is critical that the device delivers consistent outputs across different wavelengths. Our design leverages superposition theory~\cite{lin2005collinear, goodman1968introduction, perina1967superposition, kai2019performances}, in which the transformation of multi-wavelength optical fields is modeled as the superposition of independent coherent transformations at each wavelength.

\begin{figure*}[!t]
\centering
\includegraphics[scale=0.6]{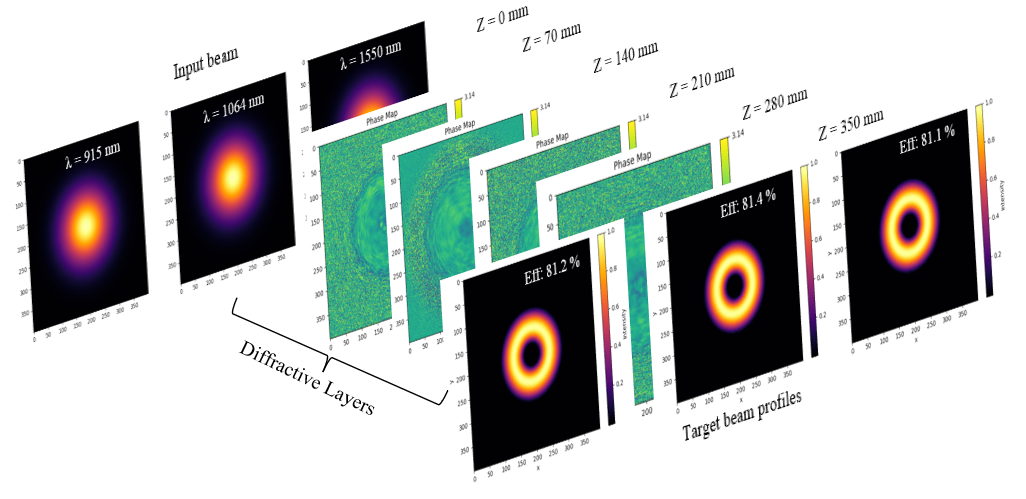}
\caption{WA-DNN schematic for generating a ring beam from input beams of different wavelengths.}
\label{fig:Wavelength_Adaptive}
\end{figure*}

Figure~\ref{fig:Wavelength_Adaptive} depicts the architecture of the Wavelength Adaptive Diffractive Neural Network (WA-DNN), where the network receives Gaussian beam profiles as input for three representative laser wavelengths. The output plane produces a predefined target intensity pattern—such as a ring-shaped beam—fixed at a specific propagation distance. Between the input and output planes, the hidden layers comprise DOEs with trainable, phase-only transmission coefficients. Each pixel in a DOE acts as a neuron, interconnecting layers via optical diffraction based on Huygens’ principle.

The network is trained using the backpropagation algorithm, and once the training is complete, the optimized phase values at each pixel of the DOEs are converted into height maps and fabricated using standard lithographic or additive manufacturing techniques. The resulting all-optical system performs parallel processing of optical information at the speed of light, without the need for electrical power during inference. By eliminating active modulators and mechanical components, the architecture provides a compact, energy-efficient, and robust platform for practical applications.

\subsection{Wave Propagation and Optimization Framework}

The DNN relies on wave propagation to connect layers. For a field propagating primarily along the $z$ -axis, the Rayleigh-Sommerfeld (RS) diffraction integral governs the transformation.

\begin{equation}
U'(x',y',z)_{\lambda_i} = \iint U(x,y,0)_{\lambda_i} \cdot h(x'-x,y'-y,z)_{\lambda_i} \, dx \, dy
\label{Field at Plane z}
\end{equation}

Here, the coordinates \( x' \) and \( y' \) correspond to the target plane. \( U'(x', y', z)_{\lambda_i} \) is the complex field at distance \( z \) along the optical axis for wavelength \( \lambda_i \). \(U(x, y, 0)_{\lambda_i} \) denotes the field in the input plane for different source wavelengths and \( h(x' - x, y' - y, z)_{\lambda_i} \) is the wavelength-dependent point spread function. 

\begin{equation}
\begin{aligned}
h(x,y,z)_{\lambda_i} &= \frac{\exp(i k_{i} r)}{r} \cdot \frac{z}{r} \left( \frac{1}{2\pi r} + \frac{1}{i \lambda_i} \right), \\
r &= \sqrt{x^2 + y^2 + z^2}
\end{aligned}
\label{Transfer function}
\end{equation}

where, z is the propagation distance from the source to either the DOE plane or the target plane. 

To efficiently evaluate the RS integral, we use the Band-Limited Angular Spectrum (BLAS) method ~\cite{matsushima2009band}, which leverages the convolution theorem by expressing the integral in terms of the Fourier transforms of \( U(x, y, 0)_{\lambda_i} \) and the wavelength-dependent point spread function \( h(x, y, z)_{\lambda_i} \).

\begin{equation}
\begin{split}
U'(x',y',z)_{\lambda_i} = 
\mathcal{F}_{u,v}^{-1}\Big\{ 
& \, \mathcal{F}_{x,y}\left\{U(x,y,0)_{\lambda_i}\right\}(u,v) \\
& \cdot H(u,v,z)_{\lambda_i}
\Big\}(x,y,z).
\end{split}
\label{Fourier_Transformed_Field}
\end{equation}

Equation~\ref{Fourier_Transformed_Field} describes the optical field after propagation to the DOE layer, where \( u \) and \( v \) denote the spectral coordinates. High-frequency components in \( H \) require fine discretization. The BLAS method employs a rectangular frequency filter and, when the sample window \( L \) is much smaller than the propagation distance \( z \), the band-limited transfer function can be approximated as shown in Equation~~\ref{BLAS_condition}.

\begin{equation}
H'_{\lambda_i}(u,v) = H_{\lambda_i}(u,v) \cdot \text{rect}\left(\frac{u}{2u_\text{lim}}\right) \cdot \text{rect}\left(\frac{v}{2v_\text{lim}}\right)
\label{BLAS_condition}
\end{equation}

with the spatial cutoff defined as:

\begin{equation}
\frac{u^2}{u_\text{lim}^2} + \frac{v^2}{v_\text{lim}^2} \leq 1, \quad u_\text{lim} = \left[ (2 \Delta u z)^2 + 1 \right]^{-1/2} \lambda_i^{-1}
\label{cutoff condition}
\end{equation}

An analogous expression holds for \( v_\text{lim} \). The field modulated by each DOE layer becomes:

\begin{equation}
U'_{\lambda_i} = T_{mn}(\Phi) \cdot U_{\lambda_i} \cdot H'_{\lambda_i}
\label{DOE modulated field}
\end{equation}

Here, \( T_{mn}(\Phi) = t_{mn} e^{i \phi_{mn}} \) encodes the phase modulation per pixel, with fixed amplitude \( t_{mn} = 1 \) to ensure power conservation. The output intensity is then:

\begin{equation}
I_{\lambda_i} = \| U'_{\lambda_i} \|^2 = \| T_{mn}(\Phi) U_{\lambda_i} H'_{\lambda_i} \|^2
\end{equation}

To enable the generation of distinct spatial profiles at different target planes for $N_\lambda$ wavelengths, we optimize the phase values of the diffractive layers accordingly. Since this is a multi-objective regression problem, we employ the mean squared error (MSE) loss function to guide the optimization of the phase values in the diffractive optical element (DOE) layers.

\begin{equation}
L_{MSE} = \frac{1}{N^2} \sum_{i=1}^{N_\lambda} \sum_{p=1}^{P} \sum_{j,k=0}^{N-1} \left( I'^{\text{Target},(Z_p)}_{\lambda_{i,jk}} - I'^{\text{Generated},(Z_p)}_{\lambda_{i,jk}} \right)^2
\end{equation}

\noindent\text{for each} $\lambda_i = 1, \dots, N_\lambda$.

Here, $N$ is the dimension of both the target and the generated beam matrices. The indices $j$ and $k$ run over the spatial dimensions of the sample window $L$ for each wavelength. $I'^{\text{Target},(Z_p)}_{\lambda_{i,jk}}$ is the target value at spatial location $(j,k)$ for the $i$-th wavelength at the $p$-th target plane located at distance $z_p$. Similarly, $I'^{\text{Generated},(Z_p)}_{\lambda_{i,jk}}$ is the generated intensity at the same spatial location $(j,k)$ for the $i$-th wavelength, also evaluated at the $p$-th target plane.

The ADAM optimizer was chosen for training the DNNs due to its adaptability, enabled by an adaptive learning rate that dynamically adjusts the step size during training, as well as its memory efficiency. In this work, a learning rate of $1\times10^{-2}$ was used, and the network was trained for 5000 iterations. The training process employed the backpropagation algorithm to optimize the phase values of the diffractive layers.

\subsection{Phase-to-Height Conversion and Implementation Details}

Once optimized, the phase masks are converted to DOE height maps using the following relation:

\begin{equation}
\Delta h = \frac{\lambda_{\text{ref}}}{2\pi \Delta n_{\lambda_{\text{ref}}}} \Phi_{mn}
\end{equation}

The reference wavelength \( \lambda_{\text{ref}} \) is taken as the average of the considered wavelengths (e.g., 915 nm, 1064 nm,  and 1550 nm). \( \Delta n \) is the refractive index contrast between the DOE material $\mathrm{SiO_2}$ and background medium which is air. Illumination with sources of different wavelengths causes the phase shift to scale inversely:

\[
\phi(x, y) \cdot \frac{\lambda_{ref}}{\lambda_{1}}, \quad \phi(x, y) \cdot \frac{\lambda_{ref}}{\lambda_{2}}, \quad \ldots, \quad \phi(x, y) \cdot \frac{\lambda_{ref}}{\lambda_{i}}.
\]

The WD-DNN framework was implemented in Python 3.8.10 using PyTorch 2.3.0. Simulations were performed on a workstation equipped with an NVIDIA RTX A2000 12GB GPU, Intel Core i9-13900K CPU, and 128 GB RAM.

\begin{figure*}[!t]
\centering
\includegraphics[scale=0.5]{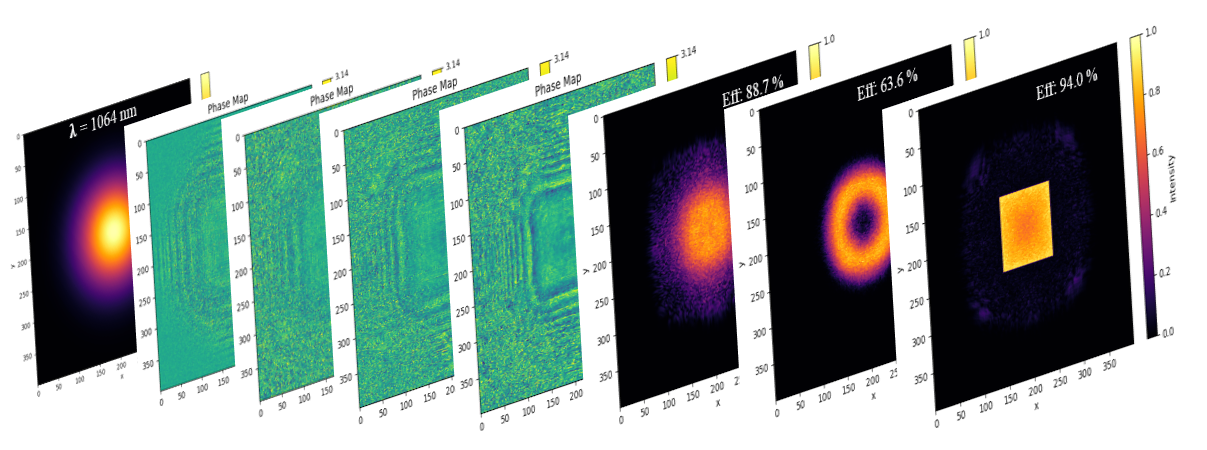}
\caption{Schematic of the DD-DNN generating Gaussian, ring, and top-hat beams at different axial distances using a single source wavelength (1064~nm).}
\label{fig:Dynamic_Beam_Schematic}
\end{figure*}

\section{DNN Functionalities}
As introduced earlier, the DNN is trained to perform beam shaping at three distinct wavelengths: 915~nm, 1064~nm, and 1550~nm. These wavelengths are commonly used in laser-based manufacturing due to their diverse interactions with metals, polymers, and composite materials. Training the network across this wavelengths enables wavelength-adaptive beam shaping within a single optical platform.

Industrial laser sources typically emit Gaussian (TEM$_{00}$) beams, which are not always optimal for specific processing tasks. Tailored beam profiles—such as top-hat, ring, or Bessel beams—can significantly improve performance and efficiency in tasks like welding, cutting, drilling, or micromachining. By learning to generate these profiles, the DNN enhances process flexibility and quality while reducing the need for multiple optical elements or setups.

Given the combined importance of wavelength versatility and beam profile customization, we train the DNN to perform a series of beam-shaping tasks under varied spectral and spatial constraints. The following sub-sections detail its capabilities and performance.
\subsection{Wavelength-Adaptive Beam Shaping}
To demonstrate the wavelength adaptability of our system, we first trained the WA-DNN to generate a ring-shaped beam profile at a fixed target plane located 350~mm from the input. This task was performed for three representative wavelengths, \textit{(i,e)} 915\,nm, 1064\,nm, and 1550\,nm—as shown in Fig.~\ref{fig:Wavelength_Adaptive}. The ability to produce the same spatial intensity pattern across different wavelengths underscores the inherent wavelength adaptability of the system. Thus proving it advantageous over existing wavelength-specific beam-shaping devices such as SLM, DOE, where the operation is limited to a specific wavelength.  

For the simulation, we modeled the input as a Gaussian beam with a 2\,mm radius, confined within a 10\,mm\,$\times$\,10\,mm window and sampled over a 400\,$\times$\,400 pixel grid. To explore the influence of network depth, we varied the number of diffractive layers between the input and target planes. The resulting efficiencies for each configuration are summarized in Table~\ref{tab:layer_efficiency}. Among these, a 4-layer architecture emerged as the optimal choice, striking a balance between beam-shaping accuracy and system complexity.

\begin{table}[!t]
\caption{Efficiency (\%) of Target Beam Generation for Different Source Wavelengths and Layer Counts}
\label{tab:layer_efficiency}
\centering
\begin{footnotesize}
\begin{tabular}{|c|c|c|c|c|}
\hline
\textbf{Wavelength} & \multicolumn{4}{|c|}{\textbf{Efficiency}} \\
\hline
 & \textbf{2 Layers} & \textbf{3 Layers} & \textbf{4 Layers} & \textbf{5 Layers} \\
\hline
915~nm   & 71.1 & 77.8 & 80.0 & 81.7 \\
1064~nm  & 73.7 & 79.4 & 81.3 & 82.0 \\
1550~nm  & 75.6 & 78.7 & 81.4 & 81.6 \\
\hline
\end{tabular}
\end{footnotesize}
\end{table}

This capability has direct implications for industrial laser processing, where materials often require different operating wavelengths due to their varied optical absorption properties. In a conventional setup, switching between materials often means switching optical components or reconfiguring the system entirely. In contrast, the WA-DNN enables material-specific processing through wavelength switching alone—without modifying the optical elements. For instance, in laser drilling applications, this flexibility allows seamless transition between metals, polymers, and composites, streamlining operations in sectors such as electronics, aerospace, and medical device manufacturing.

\subsection{Beam Profile Switching via Axial Translation}

While wavelength adaptability is crucial in laser-based materials processing—since each material exhibits different absorption characteristics—the spatial distribution of the laser beam is equally important. Different beam profiles are often required at various stages of a process, such as preheating, welding, and post-treatment. Traditionally, achieving such versatility demands either multiple optical components or active modulation systems, which add complexity and alignment challenges. Here, we demonstrate an alternative approach: a single, static optical element that enables passive switching between beam profiles simply through axial translation of the observation plane.

In this work, we propose a Depth-Dependent Diffractive Neural Network (DD-DNN) that generates distinct beam profiles at different propagation distances. As shown in Fig.~\ref{fig:Dynamic_Beam_Schematic}, when a Gaussian source of wavelength 1064~nm illuminates the optical system, it produces a Gaussian beam at 300~mm, a ring beam at 350~mm, and a top-hat beam at 400~mm. This enables users to select the beam profile of interest by merely translating either the optical system or the workpiece along the propagation axis—without modifying the device itself. The network was implemented with four diffractive layers, which provided sufficient degrees of freedom to achieve high-fidelity beam shaping.

The choice of beam profiles was deliberate, designed to meet distinct functional requirements in laser-based manufacturing. Gaussian beams, characterized by their high central intensity, are ideal for initiating localized melting or precision tasks ~\cite{semak1999laser}. Ring beams ~\cite{singh2024holistic} offer a hollow energy distribution that helps stabilize melt pool dynamics, reducing spatter and porosity—common challenges during welding. Top-hat beams ~\cite{u2014radiant}, with their uniform intensity profile, are better suited for applications requiring even energy deposition, such as surface treatments, cladding, or post-processing. By integrating all three profiles into a single passive device, the DD-DNN provides a compact, alignment-free solution for multi-stage or material-specific laser processing workflows.

Beyond static beam profile switching, the same device concept can also be extended to active process control. For instance, one may ask: what if a programmed sequence of beam profiles at a specified interval is desired for an optimal manufacturing solution? The DD-DNN architecture enables this possibility, laying the foundation for what we describe as ``dynamic beam sequencing" — a concept explored in the following section.

\begin{figure*}[!t]
\centering
\includegraphics[scale=0.6]{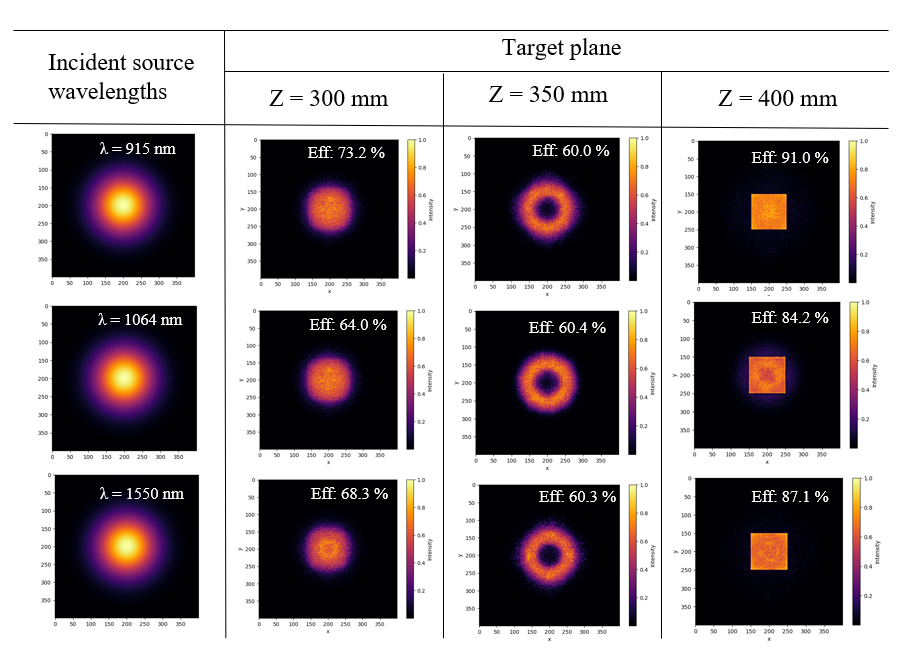}
\caption{Simulated beam profiles generated by the WD-DNN at 300~mm (Gaussian), 350~mm (ring), and 400~mm (top-hat) for three different wavelengths: 915~nm, 1064~nm, and 1550~nm.}
\label{fig:Dynamic_Beam_Profiles_at_different_distances_all_wavelengths1}
\end{figure*}

\subsection{Dynamic Beam Engineering with DNN}

While the previous sections highlighted the advantages of an optical device that is wavelength-adaptable and capable of generating different beam profiles at distinct axial planes, an even more powerful approach is to combine both functionalities—wavelength adaptability and dynamic beam profile generation—within a single system. This section explores proof-of-concept studies aimed at exploring capabilities in the context of practical manufacturing scenarios.

Modern laser-based manufacturing increasingly benefits from dynamic beam control—the ability to adapt the spatial beam profile on demand or in real time to match different stages of a process.  A particularly compelling use case is EV welding, where different beam profiles are required in sequence to optimize energy delivery, thermal gradients, and joint quality. One such example is hairpin welding ~\cite{shekel2024dynamic}, where beam shape sequencing was applied to address challenges such as porosity and spatter, which often arise from unstable melt pool dynamics when the beam crosses the gap between adjacent pins. A sequence of four localized spots was used to initiate controlled melting and bridge the gap, followed by a rapid transition--within 5~\textmu s--to a larger beam encompassing both pins. This approach ensured uniform energy delivery and improved weld integrity.

The ability to transition between these beam profiles in a controlled manner is thus highly desirable—not only for addressing the needs of individual process phases, but also for sequencing energy delivery to engineer targeted thermal and  mechanical material responses.The proposed DNN approach enables such beam sequencing in addition to supporting separate processes without the complexity of conventional architectures. Unlike traditional systems that rely on multiple components or active modulation, the DNN integrates the functionality of multiple optical elements into a single compact, static design. This paradigm shift simplifies integration, enhances reliability, and removes the need for external control mechanisms. 

Recognizing the need for greater flexibility in practical deployment, we extended the system to support multi-wavelength dynamic beam shaping. The same DD-DNN architecture was re-optimized to generate the desired beam profiles at fixed axial planes for three different industrial laser wavelengths: 915\,nm, 1064\,nm, and 1550\,nm. This extended network is referred to as the Wavelength- and Depth-Dependent Diffractive Neural Network (WD-DNN). The resulting output profiles and their corresponding diffraction efficiencies for each wavelength are shown in Fig.~\ref{fig:Dynamic_Beam_Profiles_at_different_distances_all_wavelengths1}. When integrated with a robotic arm, the optical system delivers different beam profiles at distinct axial planes, allowing for programmable sequencing—for example, directing a Gaussian beam onto the workpiece for 10~s, switching to a ring beam for 20~s, and then a top-hat beam for 30~s—simply through controlled axial translation.

This multi-wavelength, multi-profile multiplexing capability demonstrates that the WD-DNN can perform spatial sequencing of beam profiles along the propagation axis for three different wavelengths within a single, static optical design—without the need to replace or adjust the optical setup. This represents a significant advancement over conventional DOEs, which are typically narrowband and fixed-function. In contrast, the WD-DNN enables process-aware beam sequencing and material-adaptive control, which are essential for advanced applications requiring both precision and flexibility, such as laser additive manufacturing, battery welding, and microfabrication.

\section{Summary and Conclusions}

In summary, recognizing the need for on-demand beam profiling in laser-assisted manufacturing, this work demonstrates a Wavelength-Depth Dependent Diffractive Neural Network capable of performing wavelength-adaptive dynamic beam shaping at three industrially relevant wavelengths: 915 nm, 1064 nm, and 1550 nm. Initially, the network was devised to produce ring-beam profiles simultaneously at all three key wavelengths, addressing the key limitation of conventional DOEs, which are typically narrowband and designed for fixed-function beam generation. Subsequently, the network was trained to generate Gaussian, ring, and top-hat beam profiles at specific propagation distances. Apart from performing static beam shaping, the system also enables dynamic beam sequencing by merely translating either the device or the workpiece, without requiring active modulation as in the case of spatial light modulators, electro-optic modulators, acousto-optic modulators, or digital micromirror devices, to mention a few. 

The proposed architecture offers a compact, robust, and wavelength-adaptive solution suitable for industrial applications such as electric vehicle battery welding, precision cutting, additive manufacturing, surface texturing, and semiconductor processing, where different stages demand tailored beam profiles. By reducing system complexity and enhancing a wide range of material compatibility, this approach paves the way for intelligent, multifunctional laser processing systems. Overall, our results demonstrate the potential of diffractive neural networks to overcome traditional limitations in beam shaping and to enable advanced, multi-wavelength optical control for next-generation laser manufacturing technologies.

\begin{algorithm}[H]
\caption{WD-DNN Training With Multi-Wavelength and Multi-Plane Intensity Targets}
\label{alg:wddnn}
\begin{algorithmic}
\STATE 
\STATE \textbf{Input:} Target intensity profiles $I^{\text{Target}}_{\lambda_i,Z_p} \in M(N \times N, C, P)$
\STATE \textbf{Output:} Array of optimized phase masks $\phi_j$, test loss $L_{test}$
\STATE Initialize phase masks $\phi_j \in M(N \times N, \mathbb{R})$, with $\phi_{j,mn} \gets 0$
\STATE Pre-calculate transfer functions $H_{\lambda_i, Z_p} \in M(N \times N, C, P)$
\STATE Initialize initial field distribution $U_{\lambda_i} \in M(N \times N, C)$
\STATE 

\FOR{$epoch = 1$ \TO $Epochs$}
    \STATE $L_{train\_total} \gets 0$
    \FOR{$i = 1$ \TO $num\_wavelengths$}
        \STATE Initialize $U_{\lambda_i}$
        \FOR{$j = 1$ \TO $num\_layers$}
            \STATE $U_{\lambda_i} \gets \text{FFT2}^{-1}(\text{FFT2}(U_{\lambda_i}) \cdot H_{\lambda_i,j})$
            \STATE $U_{\lambda_i} \gets U_{\lambda_i} \cdot \exp(i \cdot \phi_j)$
        \ENDFOR
        \FOR{$p = 1$ \TO $num\_planes$}
            \STATE $U_{\lambda_i, Z_p} \gets \text{FFT2}^{-1}(\text{FFT2}(U_{\lambda_i}) \cdot H_{\lambda_i, Z_p})$
            \STATE $I^{\text{Generated}}_{\lambda_i, Z_p} \gets | U_{\lambda_i, Z_p} |^2$
            \STATE $L_{train\_ip} \gets L_I(I^{\text{Generated}}_{\lambda_i, Z_p}, I^{\text{Target}}_{\lambda_i, Z_p})$
            \STATE $L_{train\_total} \gets L_{train\_total} + L_{train\_ip}$
        \ENDFOR
    \ENDFOR
    \STATE $L_{train\_mean} \gets \frac{L_{train\_total}}{num\_wavelengths \times num\_planes}$
    \STATE Backpropagate gradients from $L_{train\_mean}$
    \STATE Update $\phi_j$ using ADAM optimizer
\ENDFOR
\STATE 
\RETURN $\phi_j$, $L_{test}$
\end{algorithmic}
\end{algorithm}

\section*{Acknowledgments}
The authors acknowledge UKIERI-SPARC~(3673), CEFIPRA/IFCPAR~(IFC/7148/2023), and  Extramural Research \& Intellectual Property Rights (ER\&IPR/D(R\&D)/1865) for financial support through research projects. Additionally, KN thanks the Anusandhan National Research Foundation, India, for support through the Core Research Grant (CRG/2023/008068).

\end{document}